\documentclass{iopjournal}
\usepackage{textcomp}

\begin{document}

\articletype{Paper}

\title{Effect of interatomic repulsion and quasi-degenerate states on a Kitaev-transmon qubit based on double quantum dots}

\author{Clara Palacios$^1$ and A. A. Aligia$^{1,2}$\orcid{0000-0001-9647-3926}}

\affil{$^1$Centro At\'{o}mico Bariloche and Instituto Balseiro, 8400 Bariloche, Argentina}

\affil{$^2$Instituto de Nanociencia y Nanotecnolog\'{\i}a
CNEA-CONICET, GAIDI,
Centro At\'{o}mico Bariloche, 8400 Bariloche, Argentina}

\affil{$^*$A. A. Aligia}

\email{aaligia@gmail.com}


\begin{abstract}

We  investigate the effect of interatomic Coulomb repulsion $V$
and particular states disregarded previously
on the Kitaev-transmon system proposed by Pino \textit{et al.} \cite{pino} which consists of a
Josephson junction connecting two double quantum dots (DQDs) modeled by the spinless Kitaev Hamiltonian.
For an isolated DQD, we demonstrate that a
``sweet spot'' hosting ``poor  man's Majorana'' states
persist in the presence of $V$, provided that system parameters are appropriately tuned.
For the full system, we demonstrate that, at the sweet spots of both DQDs, all eigenstates are doubly degenerate. This degeneracy arises from the existence of an operator that maps between two decoupled Hilbert subspaces.
Away from the sweet spots, the microwave spectrum becomes sensitive to the choice of initial state of the system. In our study, we consider transitions from the ground state
(which depending on the flux alternates between the above mentioned subspaces)
to all possible excited states. This scenario corresponds to a system initially in thermal equilibrium at low temperature.

\end{abstract}

\section{Introduction}

\label{intro}

In recent years, topological superconducting systems have attracted great interest
due to the presence of Majorana zero modes (MZMs) \cite{sato,kitaev,nayak,alicea,aasen,lobos,marra}, which are candidates to store and
manipulate quantum information due to their non-Abelian exchange statistics \cite{ivan}. Many studies were based on the Kitaev model for $p$-wave superconductors \cite{kitaev2}. However, due to difficulties in realizing such a model experimentally,
other systems were studied, such as nanowires with
strong spin-orbit interaction with induced superconductivity by proximity to an $s$-wave
superconductor \cite{lutchyn2010,oreg2010,wires-exp1,wires-exp2,wires-exp3,wires-exp4,casas}. After the first theoretical proposals \cite{lutchyn2010,oreg2010},
MZMs were reported experimentally \cite{wires-exp1,wires-exp2,wires-exp3,wires-exp4}.
In recent years, the phase diagram of the model \cite{diag} has been studied, along with its derived properties and implications. A few examples are in Refs. \cite{tomo,pan,qiao,mateos,feng,prada,gru,ptok}.

The system consisting of  a topological superconducting wire and a quantum dot (QD) has also been studied \cite{wires-exp4,prada,gru,ptok,ruiz,clarke,deng,ricco,souto,kenyi}.
In particular  Prada \textit{et al.} proposed that
a QD at the end of the nanowire may be used as a powerful spectroscopic tool to quantify the degree of Majorana nonlocality through a local transport measurement \cite{prada}, and this proposal has been confirmed experimentally \cite{deng}.

Artificial Kitaev chains based on QDs have been proposed to circumvent disorder and other issues \cite{leij,sau,liu,tsin,bordin23}.
However, in these systems the topological protection is lost
and the MZMs are obtained by fine tuning parameters such as
the crossed Andreev reflection (CAR) and single-electron elastic cotunneling (ECT). For this reason these modes are called
``poor man's Majoranas'' (PMMs) \cite{leij}.
The point in the space of parameters at which these
PMMs exist is called ``sweet spot'' \cite{tsin}.
The control of these
parameters has been demonstrated experimentally
in systems of two \cite{dvir,haaf} and three \cite{bordin2,bordin3} QDs.
As a consequence, systems with PMMs in double QDs (DQDs) have been theoretically
studied recently \cite{liu22,tsin24,pino,liu24,luet,sanches}.

In particular, Pino \textit{et al.} studied a qubit based on a Josephson junction connecting
two DQDs described by the spinless Kitaev model (a minimal Kitaev-transmon
qubit) \cite{pino}. They calculated the microwave spectrum as in similar systems
studied previously \cite{pers,gino,vaey,avila}.

The transmon qubits have received great attention in recent years
\cite{larsen,caspa,barge20,krin,barge22,pita,gyenis,matu}. They present several advantages over other qubit architectures, like reduced sensitivity to charge noise \cite{gyenis}, long coherence
times \cite{larsen,caspa}, and strong
coherent qubit-qubit coupling \cite{pita}. Some relevant experiments are in Refs. \cite{larsen,caspa,barge20,krin,barge22,pita}.
A related model for a spinless Kitaev chain has been investigated in recent work \cite{torres}, and the concept of Majorana-based transmons has been examined in earlier studies \cite{smith}.
The expected key advantage of the Kitaev-transmon over the ordinary transmon is its inherent topological protection and robustness of the Majorana fermions. This design suppresses its sensitivity to dielectric loss, charge noise, and quasiparticles, which are expected to significantly extend the relaxation ($T_1$) and coherence
($T_2$) times.

In this paper, we first investigate the effect of interatomic Coulomb repulsion on the PMMs in a DQD, and subsequently examine how this interaction, as well as previously disregarded states, impacts the transmon qubit and its microwave spectrum.
Theoretical studies that explicitly account for interactions in Majorana zero modes (MZMs) remain relatively scarce \cite{ruiz,kenyi,tsin,luet,ruiz,torres,Ganga,tho13,Pandey2,miao17,ger16,camja,wiec,pandey,son,samue,chine,souto25}.
Notably, several works highlight the crucial role of interatomic Coulomb repulsion
$V$ in superconducting Kitaev chains \cite{Ganga,tho13,Pandey2,miao17}. These interaction effects can be more naturally and tractably incorporated in DQDs.
Previously, theoretical works have shown the importance of
$V$  on transport through DQDs
\cite{rafa1,rafa2,ruoko,yada,heat,dar1,dar2}.

In Section \ref{2dots} we study the conditions under which PMMs exist in an interacting DQD. In Section \ref{2dqds} we discuss the eigenstates and energies of a system of two DQDs
connected by a Josephson junction. In Section \ref{trans} we study the spectrum of the whole system, the Kitaev-transmon qubit, by numerically diagonalizing the Hamiltonian for
some parameters of interest.
Section \ref{sum} contains a summary.

\section{Double quantum dot with interdot repulsion}

\label{2dots}

In this Section we analyze how interdot Coulomb repulsion $V$ affects the stability of PMMs in
DQDs and identify the parameter regime where they remain robust (the sweet spot). The model describes two spin-polarized quantum dots as in
the experiment of Dvir \textit{et al.} \cite{dvir} and therefore
the spin index can be dropped.

The Hamiltonian is given by

\begin{equation}
    H_\mathrm{DQD} =  - \sum_{i} \mu_i c_i^\dagger c_i +(\Delta  c_1 c_2- t c_1^\dagger c_2 + \mathrm{H.c.}) + V c_1^\dagger c_1 c_2^\dagger c_2,
\label{h2d}
\end{equation}
where $c_{i}^{\dagger }$ creates a spinless electron at the dot $i$. The
first term describes the chemical potential at each dot, $\Delta $ and $t$
correspond to the amplitudes of the CAR and ECT respectively and $V$ is the
interdot repulsion. We assume that the phases of $c_{i}^{\dagger }$ have
been chosen so that $\Delta >0.$

The parity of the number of particles is conserved and the Hamiltonian can
be easily diagonalized in each subspace of well defined parity. In the
subspace of even parity, the ground state $|g_{E}\rangle$ and its energy
$E_{E}$ are given by

\begin{eqnarray}
|g_{E}\rangle &=&u_{E}|0\rangle +v_{E}c_{1}^{\dagger }c_{2}^{\dagger
}|0\rangle ,\mathrm{ }E_{E}=-\mu +\frac{V}{2}-r_{E},  \nonumber \\
u_{E}^{2} &=&\frac{1}{2}+\frac{V-2\mu }{4r_{E}},\mathrm{ }v_{E}^{2}=1-u_{E}^{2},  \nonumber \\
\mu &=&\frac{\mu _{1}+\mu _{2}}{2},\mathrm{ }r_{E}=\sqrt{\left( -\mu +\frac{V}{2}\right) ^{2}
+\Delta ^{2}},  \label{even}
\end{eqnarray}
where the coefficients of the ground state $u_{E},v_{E}>0$ and
$\mu$ is the average of the chemical potentials.
Similarly, for odd parity, changing the subscript $E$ (even) for $O$ (odd),
the result is

\begin{eqnarray}
|g_{O}\rangle &=&u_{O}c_{1}^{\dagger }|0\rangle +v_{O}c_{2}^{\dagger
}|0\rangle ,\mathrm{ }E_{O}=-\mu -r_{O},  \nonumber \\
u_{O}^{2} &=&\frac{1}{2}+\frac{\delta }{2r_{O}},\mathrm{ }v_{O}^{2}=1-u_{O}^{2},  \nonumber \\
\delta &=&\frac{\mu _{1}-\mu _{2}}{2},\mathrm{ }r_{O}=\sqrt{\delta ^{2}+t^{2}},
\label{odd}
\end{eqnarray}
where $u_{O}>0$, sgn$(v_{O})=$sgn$(t)$ and $2 \delta$ is the difference
between the chemical potentials.

As discussed in Refs. \cite{tsin,luet}, a system hosting poor man’s Majoranas
(PMMs) should satisfy four conditions, outlined below. To motivate these
requirements for the unfamiliar reader, we briefly recall the defining properties
of a genuine Majorana fermion. A Majorana operator is its own antiparticle and
can be expressed as
$\gamma =f+f^{\dagger }$, where $f$ is an ordinary fermion.
Therefore, it is chargeless. We assume that $\gamma $ is a one-particle eigenoperator
of the Hamiltonian that describes the system, i.e.
$\left[ H,\gamma \right] =E_{\gamma } \gamma $. Taking the Hermitian
conjugate of this relation one obtains the same equation with the
opposite sign of
$E_{\gamma }$ implying $E_{\gamma }=0$.
Therefore, if the ground state for an even number of particles is denoted by
$|E\rangle $, then the state $|O\rangle =\gamma
|E\rangle $ is a state with an odd number of particles and the same
energy. Note that since $\gamma ^{2}=1$, $|E\rangle =\gamma |O\rangle $.
For PMMs, one seeks to replicate these key properties of Majoranas, with two additional requirements: perfect localization at a single quantum dot (the third condition, essential for braiding operations), and the presence of a finite excitation gap, ensuring robustness of the system against disorder and small perturbations
(fourth condition).
If $\gamma $ is localized at one site $j$, it commutes with the occupancy
operator $c_{k}^{\dagger }c_{k}$ for $k \neq j$. Therefore, the charge
under addition of $\gamma $ is conserved at all sites
rather than merely conservation of the total charge.

A local Majorana fermion at the site $j$ of the DQD has the form
$\tilde{\gamma}_{j}=\tilde{c}_{j}+\tilde{c}_{j}^{\dagger }$, with
$\tilde{c}_{j}=e^{i\alpha }c_{j}$ for some convenient phase $\alpha $.
We denote by $k$ the site opposite to $j.$ With this notation in place, it is natural to formulate the four conditions as follows

1) The states described by Eqs. (\ref{even}) and (\ref{odd}) should be
degenerate: $E_{O}=E_{E}$.

2) The change of the charge at each site should be zero, implying

\begin{equation}
\Delta Q_{i}=\langle g_{E}|c_{i}^{\dagger }c_{i}|g_{E}\rangle -\langle
g_{O}|c_{i}^{\dagger }c_{i}|g_{O}\rangle =0.  \label{dq}
\end{equation}

3) The PMMs should be perfectly localized at each dot

\begin{equation}
|\langle g_{E}|\tilde{\gamma}_{j}|g_{O}\rangle |=1,\:
|\langle g_{E}|\tilde{\gamma}_{k}|g_{O}\rangle |=0  \label{local}
\end{equation}

4) The PMMs should be separated from the excited states from a finite gap.

From Eqs. (\ref{even}),(\ref{odd}) and (\ref{dq}) one obtains

\begin{equation}
\Delta Q_{1} =\frac{2\mu -V}{4r_{E}}-\frac{\delta }{2r_{O}}=0, \:
\Delta Q_{2} =\frac{2\mu -V}{4r_{E}}+\frac{\delta }{2r_{O}}=0.  \label{dq2}
\end{equation}

This implies $\delta =0$, $\mu =V/2$, or $\mu _{1}=\mu _{2}=V/2$. These
requirements for the sweet spot can be obtained by tuning the gate voltages.
The remaining requirement comes from the first condition, which is satisfied
if $\Delta =|t|+V/2$.
The same equations were obtained in recent work; however, the four conditions required for the realization of PMMs were not analyzed in detail there \cite{samue}.

Under these requirements, it can be easily checked that depending on the
sign of $t$, either $|\langle g_{E}|c_{1}^{\dagger }+c_{1}|g_{O}\rangle
|=|\langle g_{E}|i(c_{2}-c_{2}^{\dagger })|g_{O}\rangle |=1$ and $|\langle
g_{E}|c_{2}^{\dagger }+c_{2}|g_{O}\rangle |=|\langle
g_{E}|i(c_{1}-c_{1}^{\dagger })|g_{O}\rangle |=0$ or the same permuting 1
and 0, satisfying the third condition. For example $|\langle
g_{E}|c_{1}^{\dagger }+c_{1}|g_{O}\rangle |=u_{E}u_{O}+v_{E}v_{O}=[1+$sgn$(t)]/2$.
Finally, the double degenerate ground state is separated by a gap $2|t|$ from the next excited state.

In summary, in the presence of an interdot interaction $V$,
a sweet spot analogous to the case $V=0$
can still be achieved, leading to well-defined PMMs.
This sweet spot is characterized by the conditions
$\mu _{1}=\mu _{2}=V/2$ and $\Delta =|t|+V/2$.
The first condition can be readily realized experimentally by tuning the gate voltages, whereas the second depends explicitly on the value of
$V$ and may become difficult to satisfy for large $V$. Experimentally, the precise value of $V$ is generally unknown and may be challenging to estimate. From a theoretical perspective, however, interdot interactions are expected to play an important role in semiconductors with induced superconductivity and have therefore been included in several relevant studies \cite{Ganga,tho13,Pandey2,miao17}.

\section{Two DQDs connected by a Josephson junction}

\label{2dqds}

\begin{figure}[th]
    \centering
    \includegraphics[width=0.9\linewidth]{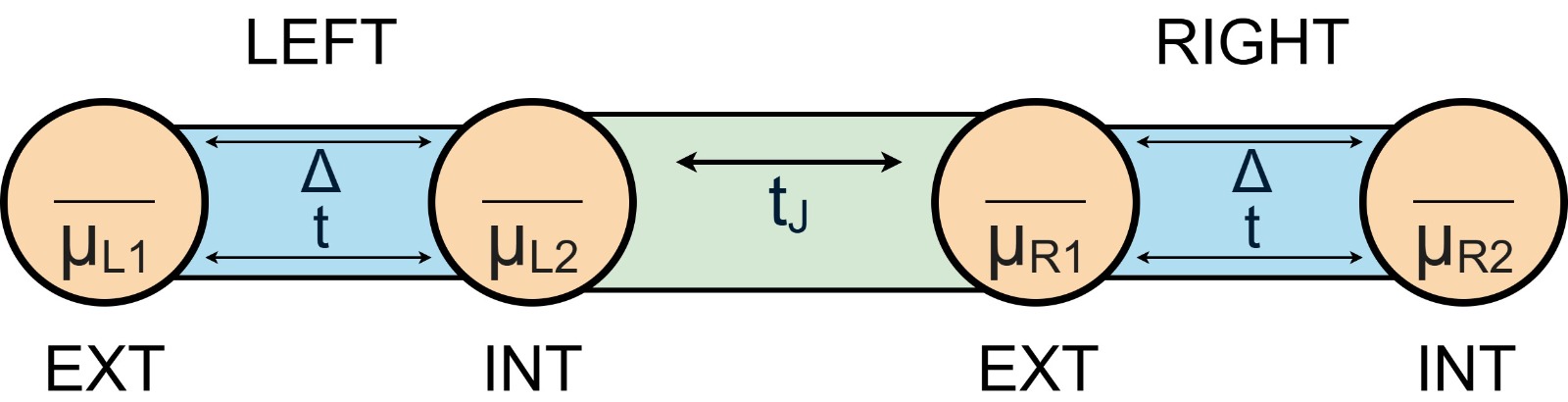}
    \caption{Scheme of the LEFT and RIGHT DQDs with CAR (ECT) amplitudes
    $\Delta $ ($t$) coupled through a Josephson junction with energy $t_J$.
    The chemical potential of each dot is denoted by $\mu_{\nu j}$.}
    \label{2dqd}
\end{figure}

In this Section, we consider two DQDs described by Eq. (\ref{h2d}) adding a
label $L$ $(R)$ for the DQD at the left (right) and coupled by the term

\begin{equation}
H_{J}=-t_{J}e^{i\phi /2}c_{L2}^{\dagger }c_{R1}+\mathrm{H.c.,}  \label{hj}
\end{equation}
so that the total Hamiltonian is

\begin{equation}
H_{\mathrm{4D}}=H_{\mathrm{DQD}}^{L}+H_{\mathrm{DQD}}^{R}+H_{J}.  \label{h4d}
\end{equation}
A scheme of the system is represented in Fig. \ref{2dqd}. For simplicity
we assume equal parameters for both DQDs, and that they are at the sweet
spot except that $\Delta $ can be different from $t+V/2$. We also take $t>0$
(the sign of $t$ can be changed by the substitution $c_{L1}\rightarrow
-c_{L1}$, $c_{R2}\rightarrow -c_{R2}$). In this case, the ground state for
$t_{J}=0$ is either $|g_{E}\rangle $ for both DQDs (a state which we denote
by $|ee\rangle $) if $\Delta >t+V/2$ or $|g_{O}\rangle $ for both DQDs
($|oo\rangle $) if $\Delta < t+V/2$.
Both states have an even total number of particles. In more general
situations, the ground state can have a total odd number of particles.

The term $H_{J}$ mixes $|ee\rangle $ with $|oo\rangle$.
Calculating
the matrix element between these two states leads to the matrix

\begin{equation}
H_{\mathrm{4D}}(\phi )=\left(
\begin{array}{cc}
E_{ee} & -\frac{t_{J}}{2}\cos {\left( \frac{\phi }{2}\right) } \\
-\frac{t_{J}}{2}\cos {\left( \frac{\phi }{2}\right) } & E_{oo}
\end{array}
\right) ,  \label{matriz}
\end{equation}
where $E_{ee}=2E_{E}$ and $E_{oo}=2E_{O}$.
This matrix has the same form as that considered by Pino \textit{et
al.} for $V=0$ \cite{pino}. The difference lies in the values
of $E_{E}$, $E_{O}$ and the conditions for the sweet spots.

\section{The Kitaev-transmon qubit}

\label{trans}

\begin{figure}[bh]
    \centering
    \includegraphics[width=0.9\linewidth]{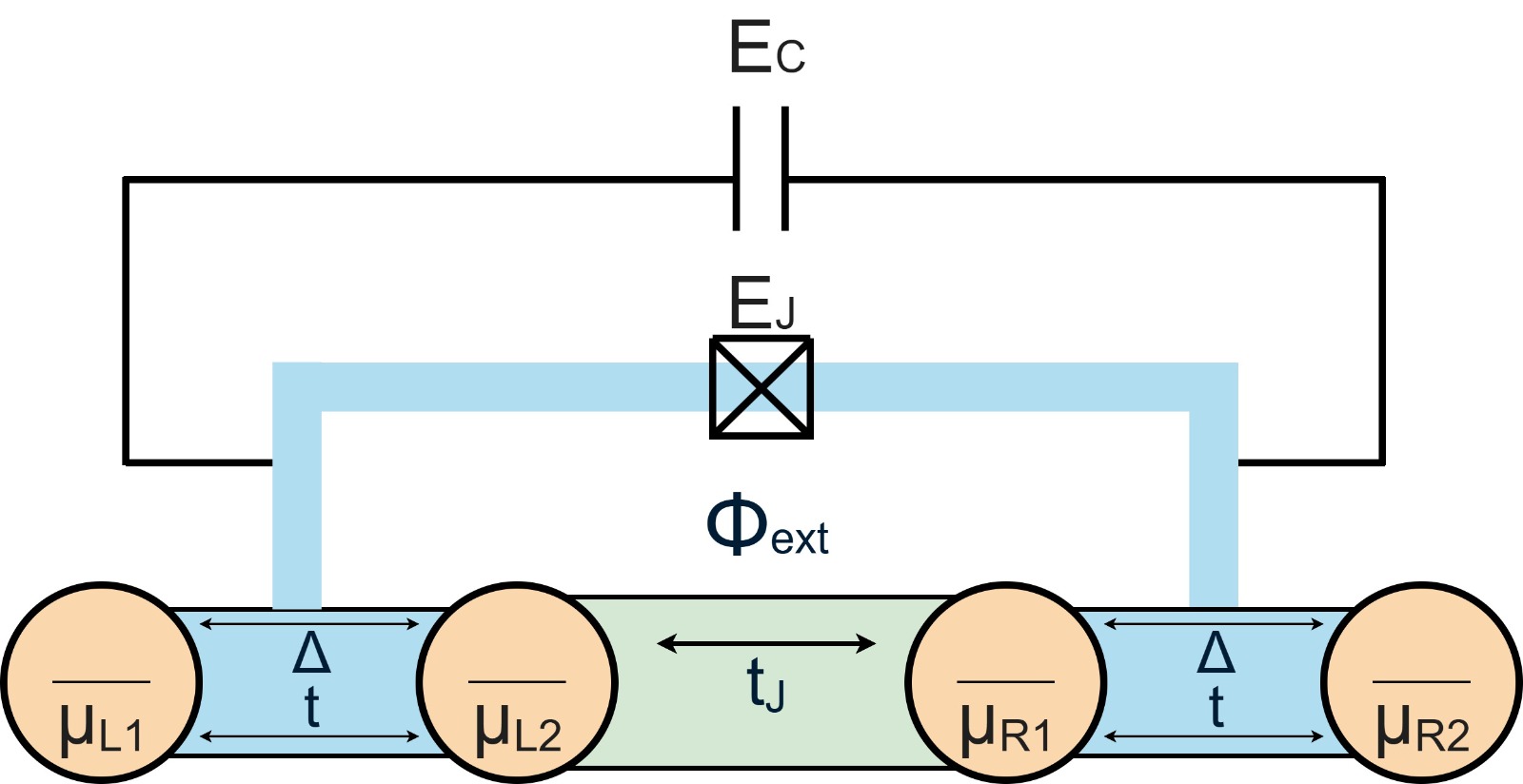}
    \caption{Scheme of the Kitaev-transmon qubit. A capacitance with energy
    $E_C$ and a Josephson junction with an energy $E_J$ are added to the two DQDs
    closing the circuit.}
    \label{transmon}
\end{figure}

In this Section we add to the previous setup a superconducting wire
containing another Josephson junction with characteristic energy $E_J$ and
together with a capacitive shunt with charging energy $E_C$, connecting the superconducting parts
of the double quantum dots closing the circuit.
A scheme of the system is shown in Fig. \ref{transmon}.
The total Hamiltonian is
\cite{pino}
\begin{eqnarray}
H &=&H_{\mathrm{4D}}(\hat{\phi}-\phi _{ext})+H_{C}+H_{J}^{\prime },  \nonumber \\
H_{C} &=&4E_{C}(\hat{m}-m_{g})^{2},\mathrm{ }H_{J}^{\prime }=-E_{J}\cos (\hat{\phi}).  \label{ham}
\end{eqnarray}
In contrast to the previous section, here the phase  $\hat{\phi}$ is treated as an {\em operator} conjugate to the operator $\hat{m}$, which is defined such that the number of Cooper pairs on the left and right sides of the junction are $m_L-\hat{m}$ and $m_R+\hat{m}$. In this framework, the expectation value of
$\hat{m}$ increases or decreases by 1 when a Cooper pair tunnels across the junction. The eigenstate of $\hat{m}$ with eigenvalue zero corresponds to charge neutrality on both sides. Consequently, $\hat{m}=-i\partial /\partial \hat{\phi}$
\cite{ferrel,newr}.
The quantity $m_{g}$
accounts for a charge imbalance controlled by an applied gate voltage.
The phase $\phi_{ext}$ is related to the external magnetic flux $\Phi $ through the circuit
by $\Phi =\Phi _{0}\phi _{ext}/(2\pi )$, where $\Phi _{0}=h/(2e)$ is the
flux quantum.

For convenience we will use the operator $\hat{n}=2\hat{m}$, and define
$\hat{\theta}=\hat{\phi}/2$ (so that $\hat{n}$ changes by 1 when
a single particle crosses the junction). Note
that the commutation relation $[\hat{\phi},\hat{m}]=[\hat{\theta},\hat{n}]=i$
retains the same form with the new variables. The low-energy states of the
system now have the form $|een\rangle $ ($|oon\rangle )$ if each DQD has
an even (odd) number of particles, and $n$
is the eigenvalue of $\hat{n}$.
Now using $e^{ib\hat{\theta}}|n\rangle
=|n+b\rangle $ (this can be demonstrated in a similar way as for the
traditional translation operator in quantum mechanics), it is clear that in
the new basis

\begin{equation}
H_{J}^{\prime }=-\frac{E_{J}}{2}\sum_{n=-\infty }^{\infty
}\sum_{p}|ppn+2\rangle \langle ppn|+\mathrm{H.c.},  \label{hpj}
\end{equation}
with $p=e$ or $o$ indicating parity,
which represents a jump of a Cooper pair from one side to the other of the
junction \cite{ferrel,newr}.

In a similar way, for the off-diagonal part of $H_{\mathrm{4D}}(\hat{\phi}-\phi
_{ext})$

\begin{eqnarray}
&&2\cos (\hat{\theta}-\theta _{ext}) =\sum_{n=-\infty }^{\infty }[(\cos
(\theta _{ext})-i\sin (\theta _{ext})) (|oon+1\rangle \langle een|+|een+1\rangle \langle oon|)+\mathrm{H.c.}]
\label{od}
\end{eqnarray}

From the structure of the Hamiltonian, it is clear that the Hilbert space can be
divided in two decoupled subspaces: a) states
$|een_1\rangle$ and $|oon_2\rangle$ with $n_1$ even and $n_2$ odd and b)
the same with $n_1$ odd and $n_2$ even. Rather surprisingly, for
$E_{oo}=E_{ee}$ (a situation which includes the sweet spots of both DQDs)
and $\phi _{ext}=0$, the eigenvalues of both subspaces are identical.
The demonstration is included in the Appendix.
Our results indicate that only the first subspace (in which the difference
of particles between both sides of the junction is even) was considered
in Ref. \cite{pino}. The authors have chosen one subspace because
they were interested in the properties of the qubit, which are similar
in the different subspaces (including $|eon \rangle$ and $|oen\rangle$)
\cite{note}.

Due to the above mentioned degeneracy, our results
for the eigenergies and transitions are very similar to those of
Ref. \cite{pino} for $E_{oo}-E_{ee}=\phi _{ext}=0$ (considering that
now $E_{ee}$ includes the effects of $V$), but are richer
for other parameters.

\begin{figure}[th]
    \centering
    \includegraphics[width=0.9\linewidth]{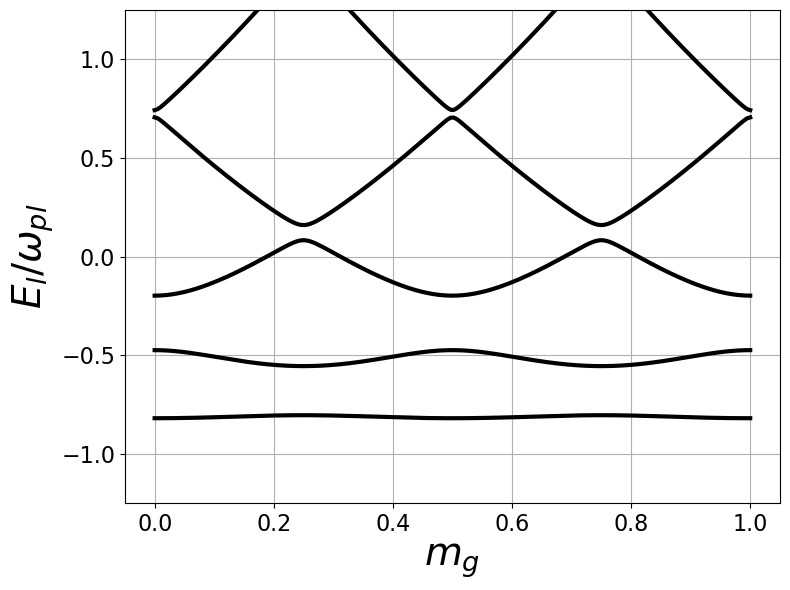}
    \caption{Energies in units of the plasma frequency
    $\omega_{pl}=\sqrt{8 E_J E_C}$, as a function of the offset charge for $\phi _{ext}=0$
    and $E_{ee}=E_{oo}$.
    Parameters are $t_J=E_J=E_c=1$, $\mu_1=\mu_2=V/2$,
$E_O=-t-V/2=E_{oo}/2=1$, and $E_E=-\Delta =E_{ee}/2$.}
    \label{sweet}
\end{figure}

We have numerically diagonalized $H$ truncating the basis to $|n_i|<16$, which
is enough to obtain accurate results. 
We take parameters similar to Ref. \cite{pino}
but adapted to the fact that $V$ can be different from zero: $t_J=E_J=E_c=1$, $\mu_1=\mu_2=V/2$,
$E_O=-t-V/2=E_{oo}/2=1$, and vary $E_E=-\Delta =E_{ee}/2$.

\begin{figure}[th]
    \centering
    \includegraphics[width=0.7\linewidth]{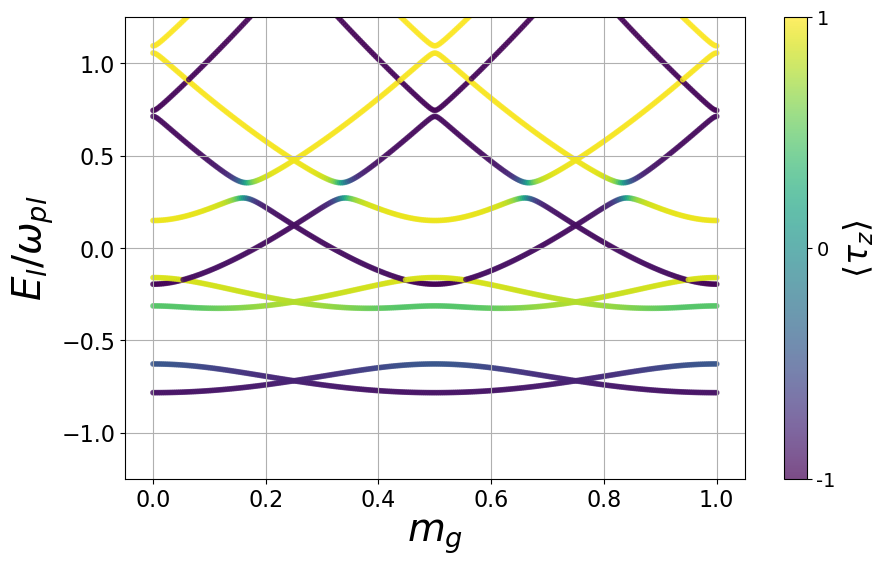}
    \includegraphics[width=0.7\linewidth]{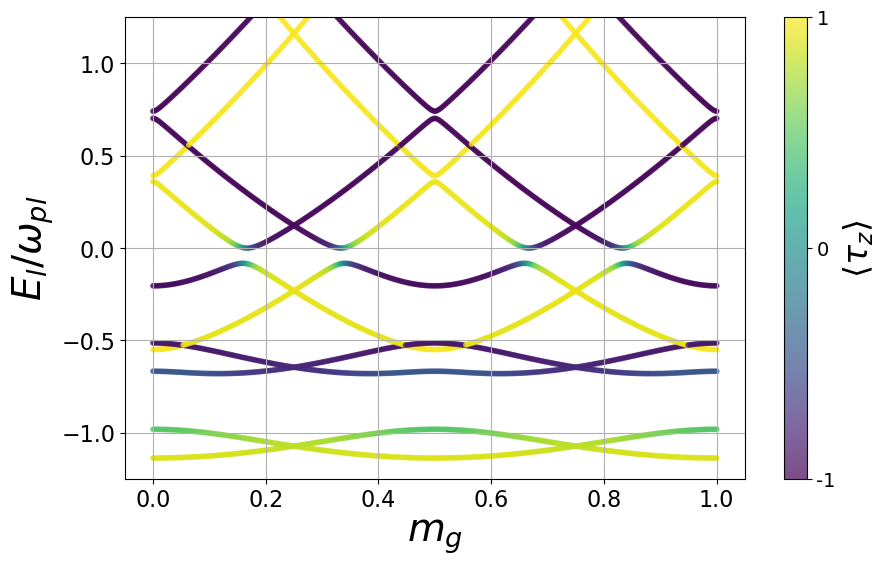}
    \caption{Energies as a function of the offset charge for $\phi _{ext}=0$ and $E_{ee}=E_{oo}/2$ (top) and $E_{ee}=3E_{oo}/2$ (bottom).
Other parameters as in Fig. \ref{sweet}.}`
    \label{ene2}
\end{figure}

The resulting energies for the 
case $E_{oo}-E_{ee}=\phi _{ext}=0$ as a function of $m_g$ are represented
in Fig. \ref{sweet}. These energies coincide with those published in Ref. \cite{pino}.
However, for $E_{ee} \neq E_{oo}$, the degeneracy between the eigenstates of both subspaces mentioned above is broken and the total energy spectrum becomes richer.
Working for parameters out of the fine-tuning condition
required for the sweet spots is more realistic and braiding can still be achieved \cite{Huang}.
The energy spectrum for
two of such cases is represented in  Fig. \ref{ene2}, where also
the expectation value of
\begin{equation}
    \tau_z = \sum_{n} (|{ee, n}\rangle \langle {ee,n}| - |{oo, n}\rangle \langle {oo,n}|)
\label{tauz}
\end{equation}
is shown.

As expected, since both energies are negative, when $E_{ee}=E_{oo}/2$, the
eigenstates of lowest energy are dominated by states $|oon \rangle$ with
$n=0$ or 1, and then
$\langle \tau_z \rangle <0$. Instead, for $E_{ee}=3E_{oo}/2$, the
low-energy eigenstates have a large weight of $|een \rangle$ and $\langle \tau_z \rangle >0$.
In both cases for $m_g=0.25$ (half a particle offset), there is a crossing of low-energy levels which belong to the different subspaces of the Hilbert space mentioned above.
For $|E_{ee} |< |E_{oo}|$, the ground state is dominated by $|oo0 \rangle$ for $m_g=0$
and by $|oo1 \rangle$ for $m_g=0.5$ as expected. Instead for
$|E_{ee}| > |E_{oo}|$,
the $|een \rangle$ states have the largest weight on the ground state.
Similar crossings take place for $m_g=0.75$.
Due to the particular parameters chosen, the energies for the case
$E_{ee}=3E_{oo}/2$ coincide with those for $E_{ee}=E_{oo}/2$, up to a global energy shift of -1 and an exchange of the corresponding Hilbert subspaces.

\begin{figure}[th]
    \centering
    \includegraphics[width=0.7\linewidth]{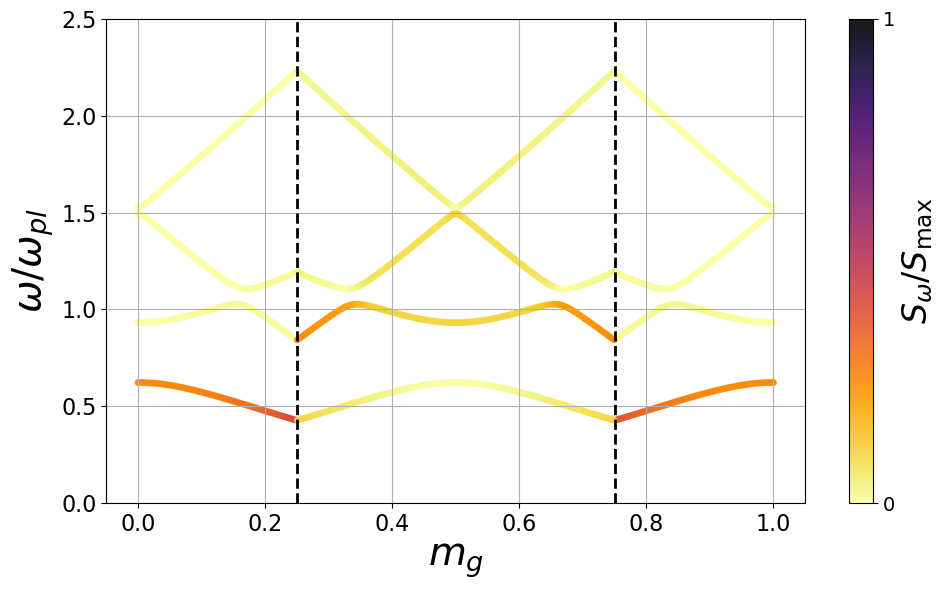}
    \includegraphics[width=0.7\linewidth]{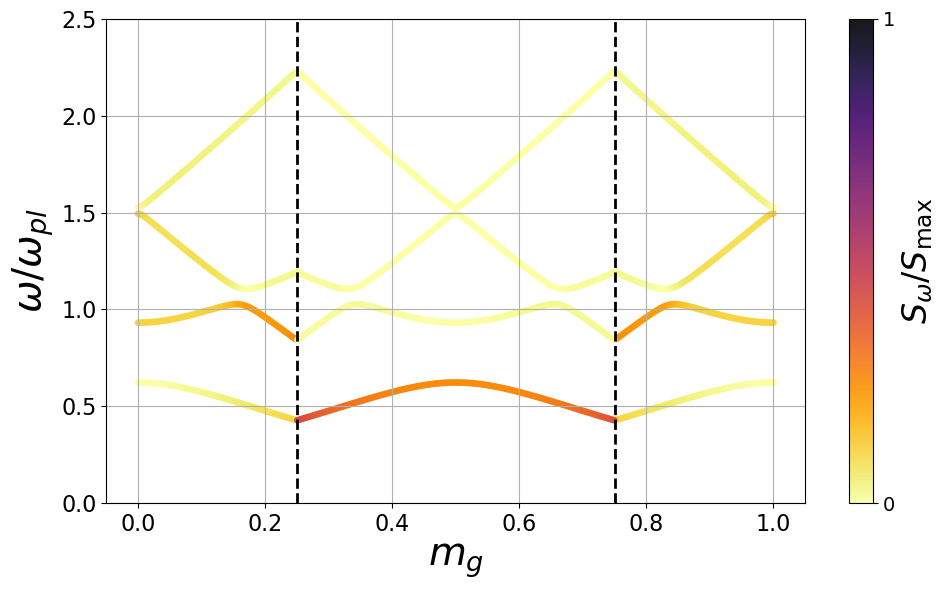}
    \caption{Microwave spectrum for $\phi _{ext}=0$ and $E_{ee}=E_{oo}/2$ (top) and $E_{ee}=3E_{oo}/2$ (bottom).
    The vertical lines indicate crossings of the ground state.
    Other parameters as in Fig. \ref{sweet}.}`
    \label{mw}
\end{figure}

From the eigenstates $|l \rangle$ and energies $E_l$ of the system, it is possible to
calculate the microwave spectrum, which at zero temperature and in linear response
is proportional to \cite{pino}
\begin{equation}
    S(\omega) = \sum_{l} |\langle l|\hat{n}|g \rangle|^2 \delta(\omega+E_g-E_l),
\label{micro}
\end{equation}
where $|g \rangle$, $E_g$ are the ground state and its energy.

For parameters corresponding to the double degeneracy of all levels,
$E_{oo}-E_{ee}=\phi _{ext}=0$, our results are similar to those presented
in Fig. 3 (e) of Ref. \cite{pino}. Instead, in the general case, the crossing
of levels mentioned above changes the ground state and affects the spectrum.
In Fig. \ref{mw} we show the frequencies and intensities of the microwave spectrum
for the same parameters as in Fig. \ref{ene2}. While our results are qualitatively
similar to those of  Ref. \cite{pino}, there are differences in the regions
$m_g<0.25$ and  $m_g > 0.75$ for $E_{ee}=E_{oo}/2$, and in the interval $0.25<m_g<0.75$
for $E_{ee}=3E_{oo}/2$ because of the change in the ground state.

Note that the transition energies coincide for both cases as a direct consequence of the particular mapping of the eigenvalues of both Hilbert subspaces mentioned above.
Due to the level crossings at $m_g=0.25$ and $m_g=0.75$, the intensities
jump at these points.

We caution the reader that quasiparticle poisoning may induce switching between the two subspaces involved at these crossing points.
As a result, the observed spectrum could appear as a superposition of the spectra associated with each
subspace \cite{pers}.

\section{Summary}

\label{sum}

We have demonstrated that sweet spots with well defined PMMs, satisfying
the four conditions for their existence, persist in DQDs described
by the Kitaev model in presence of interatomic Coulomb repulsion $V$.
Achieving this requires shifting the chemical potential of both dots and increasing the CAR amplitude $\Delta$ (or decreasing the ECT amplitude if possible) by $V/2$. This might be difficult for large $V$ indicating
that interatomic repulsion is detrimental for achieving the sweet spots
but does not necessarily render them impossible.

These results provide a simple extension of the results of Pino \textit{et al.} \cite{pino}
for the Kitaev transmon containing two DQDs, two Josephson junctions and a charge reservoir (see Fig. \ref{transmon}) by incorporating Coulomb repulsion through simple parameter renormalization (replacing $t$ by $t+V/2$ and setting $\mu_1=\mu_2=V/2$).

Outside of the sweet spots, there is an influence of states previously disregarded,
specifically, $|een\rangle $ with $n$ even and $|oon\rangle$ with $n$ odd.
We have calculated the microwave spectrum for some selected cases
which display the effect of these states. This spectrum
might be used to detect the sweet spots.

At the sweet spots of both DQDs (assumed identical), all states are
doubly degenerate, due to a particular symmetry  of the Hamiltonian, as
explined in the Appendix.

Given the broad interest in these systems, our findings offer important insights for future studies.

\section*{Acknowledgments}

We thank Daniel Dominguez,
Karen Hallberg, D. M. Pino and particularly Leandro Tosi
for useful discussions.
C. P. has a scholarship of Instituto Balseiro.
A. A. A. acknowledges financial support provided by PICT 2020A-03661 of the Agencia I+D+i, Argentina.

\appendix
\section{Double degeneracy of the eigenstates for $E_{oo}=E_{ee}$ and
$\phi _{ext}=0$}
\label{app}

The degeneracy is due to a particular symmetry of the Hamiltonian for these
parameters. To show this fact we define the operator

\begin{equation}
\hat{O}=\sum_{n}(-1)^{n}(|oon\rangle \langle een|-|een\rangle \langle oon|).
\label{ope}
\end{equation}
It is easy to see that $\hat{O}^{2}=-1$. We will show that it commutes with
the Hamiltonian. First we consider the term Eq. (\ref{od}) applied to a
state  $|oon\rangle ,$

\begin{eqnarray}
2\hat{O}\cos (\hat{\theta})|oon\rangle  &=&\hat{O}(|een+1\rangle
+|een-1\rangle ) = -(-1)^{n}(|oon+1\rangle +|oon-1\rangle ),  \nonumber \\
2\cos (\hat{\theta})\hat{O}|oon\rangle  &=&-2\cos (\hat{\theta})(-1)^{n}|een\rangle   =-(-1)^{n}(|oon+1\rangle +|oon-1\rangle ).  \label{cnd}
\end{eqnarray}
A similar result is obtained appying the oparators to $|een\rangle $.
Therefore $[\hat{O},\cos (\hat{\theta})]=0$. The remaining terms of the
Hamiltonian for $E_{oo}=E_{ee}$ consist of a sum of terms of the form
$H_{ij}=|eei\rangle \langle eej|+|ooi\rangle \langle eej|$. After a simple
algebra one obtains

\begin{equation}
\lbrack \hat{O},H_{ij}]=\left[ (-1)^{i}-(-1)^{j}\right] (|ooi\rangle \langle
eej|-|eei\rangle \langle ooj|).  \label{conm}
\end{equation}
For the diagonal terms $j=i$, while for $H_{J}^{\prime }$ [see Eq. (\ref{hpj})]  $j=i\pm 2$. Therefore the first factor cancels. This demonstrates that
$[\hat{O},H]=0$.

If $|a\rangle $ is an eigenstate of $H$ with eigenvalue $E_{a}$ ($H|a\rangle=E_{a}|a\rangle $), then $|b\rangle =\hat{O}|a\rangle $ is also an eigenstate with the same eigenvalue ($H|b\rangle =H\hat{O}|a\rangle =\hat{O}H|a\rangle =E_{a}|b\rangle $). Since the matrix of the Hamiltonian is real, and the eigenvectors can be chosen real, if $|b\rangle $ and $|a\rangle $ represent the same state, one should have $|b\rangle =\pm |a\rangle $. However, this leads to $\hat{O}^{2}|a\rangle =1$ contradicting the fact that $\hat{O}^{2}=-1$. Therefore, all eigenvectors are doubly degenerate.


\begin{thebibliography}{99}
\bibitem{sato} M. Sato and Y. Ando, Topological superconductors: a review,
Rep. Prog. Phys. \textbf{80}, 076501 (2017).

\bibitem{kitaev} A. Kitaev, Fault-tolerant quantum computation by anyons,
Ann. Phys. (N.Y.) \textbf{303}, 2 (2003).

\bibitem{nayak} C. Nayak, S. H.Simon, A. Stern, M. Freedman, and S. Das
Sarma, Non-Abelian anyons and topological quantum computation, Rev. Mod.
Phys. \textbf{80}, 1083 (2008).

\bibitem{alicea} J. Alicea, New directions in the pursuit of Majorana
fermions in solid state systems, Rep. Prog. Phys. \textbf{75}, 076501,
(2012).

\bibitem{lobos} X-J. Liu and A. M. Lobos, Manipulating Majorana fermions in
quantum nanowires with broken inversion symmetry, 
Phys. Rev. B \textbf{87}, 060504(R) (2013).

\bibitem{marra} P. Marra,
Majorana nanowires for topological quantum computation,
Journal of Applied Physics \textbf{132}, 231101 (2022).

\bibitem{aasen} D. Aasen, M. Hell, R. V. Mishmash, A. Higginbotham, J. Danon, M. Leijnse, T. S. Jespersen, J. A. Folk, C. M. Marcus, K. Flensberg, and J. Alicea, 
Milestones Toward Majorana-Based Quantum Computing, 
Phys. Rev. X \textbf{6}, 031016 (2016).

\bibitem{ivan} D. A. Ivanov,
Non-Abelian Statistics of Half-Quantum
Vortices in p-Wave Superconductors, Phys. Rev. Lett.
\textbf{86}, 268 (2001)


\bibitem{kitaev2} A. Y. Kitaev, 
Unpaired Majorana fermions in quantum wires,
Phys. Usp. \textbf{44}, 131 (2001).

\bibitem{lutchyn2010} R. M. Lutchyn, J. Sau, and S. Das Sarma, Majorana
Fermions and a Topological Phase Transition in Semiconductor-Superconductor
Heterostructures, Phys. Rev. Lett. \textbf{105} 077001 (2010).

\bibitem{oreg2010} Y. Oreg, G. Refael, and F. von Oppen, Helical Liquids and
Majorana Bound States in Quantum Wires, Phys. Rev. Lett. \textbf{105} 177002 (2010).

\bibitem{wires-exp1} V. Mourik, K. Zuo, S. M. Frolov, S. R. Plissard, E. P.
a. M. Bakkers, and L. P. Kouwenhoven, Signatures of Majorana fermions in in
hybrid superconductor-semiconductor nanowire devices, Science \textbf{336},
1003 (2012).

\bibitem{wires-exp2} A. Das, Y. Ronen, Y. Most, Y. Oreg, M. Heiblum, and H.
Shtrikman, Zero-bias peaks and splitting in an Al-InAs nanowire topological
superconductor as a signature of Majorana fermions, Nat. Phys. \textbf{8},
887 (2012).

\bibitem{wires-exp3} S. M. Albrecht, A. P. Higginbotham, M. Madsen, F.
Kuemmeth, T. S. Jespersen, J. Nyg, P. Krogstrup, and C. M. Marcus,
Exponential protection of zero modes in Majorana islands, Nature \textbf{531}, 206 (2016).

\bibitem{wires-exp4} M. Deng, S. Vaitiek\'{e}nas, E. Hansen, J. Danon, M.
Leijnse, K. Flensberg, J. Nyg\aa rd, P. Krogstrup, and C. Marcus, Majorana
bound state in a coupled quantum-dot hybrid-nanowire system, Science \textbf{354}, 1557 (2016).

\bibitem{casas} Oscar E. Casas, Liliana Arrachea, William J. Herrera, and Alfredo Levy Yeyati,
Proximity induced time-reversal topological superconductivity in
Bi$_2$Se$_3$ films without phase tuning,
Phys. Rev. B \textbf{99}, 161301(R) (2019).

\bibitem{diag} D. P\'erez Daroca and A. A. Aligia,
Phase diagram of a model for topological superconducting wires,
Phys. Rev. B \textbf{104}, 115125 (2021).

\bibitem{tomo} A. A. Aligia, D. P\'{e}rez Daroca, and L. Arrachea,
Tomography of Zero-Energy End Modes in Topological Superconducting Wires,
Phys. Rev. Lett. \textbf{125}, 256801 (2020).

\bibitem{pan} Xiao-Hong Pan, Xun-Jiang Luo, Jin-Hua Gao, and Xin Liu,
Detecting and braiding higher-order Majorana corner states through their spin degree of freedom,
Physical Review B, \textbf{105}, 195106 (2022).

\bibitem{qiao}
Guo-Jian Qiao, Xin Yue, and  C.P. Sun,
Dressed Majorana Fermion in a Hybrid Nanowire,
Phys. Rev. Lett. 133, 266605 (2024).

\bibitem{mateos} Juan Herrera Mateos, Leandro Tosi, Alessandro Braggio, Fabio Taddei, and Liliana Arrachea,
Nonlocal thermoelectricity in quantum wires as a signature of Bogoliubov-Fermi points,
Phys. Rev. B \textbf{110}, 075415 (2024).

\bibitem{feng} Junya Feng, Henry F. Legg, Mahasweta Bagchi, Daniel Loss, Jelena Klinovaja and Yoichi Ando,
Long-range crossed Andreev reflection in a topological insulator nanowire proximitized by a superconductor,
Nat. Phys. \textbf{21}, 708 (2025).

\bibitem{prada} E. Prada, R. Aguado, and P. San-Jose, 
Measuring Majorana nonlocality and spin structure with a quantum dot, 
Phys. Rev. B \textbf{96}, 085418 (2017).

\bibitem{gru} L. Gru\~{n}eiro, M. Alvarado, A. Levy Yeyati,  and L. Arrachea,
Transport features of a topological superconducting nanowire with a quantum dot: Conductance and noise,
Phys. Rev. B \textbf{108}, 045418 (2023).

\bibitem{ptok} A. Ptok, A. Kobialka, and T. Doma\'{n}ski, 
Controlling the bound states in a quantum-dot hybrid nanowire, 
Phys. Rev. B \textbf{96}, 195430 (2017).

\bibitem{ruiz} D. A. Ruiz-Tijerina,  E. Vernek,
Luis G. G. V. Dias da Silva, and J. C. Egues,
Interaction effects on a Majorana zero mode leaking into a quantum dot,
Phys. Rev. B \textbf{91}, 115435 (2015).

\bibitem{clarke} D. J. Clarke, 
Experimentally accessible topological quality
factor for wires with zero energy modes, 
Phys. Rev. B \textbf{96}, 201109(R) (2017).

\bibitem{deng} M.-T. Deng, S. Vaitiek\'{e}nas, E. Prada, P. San-Jose, J. Nyg\aa{}rd, P. Krogstrup, R. Aguado, and C. M. Marcus, Nonlocality of Majorana modes in hybrid nanowires,
Phys. Rev. B \textbf{98}, 085125 (2018).

\bibitem{ricco} L. S. Ricco, Y. Marques, J. E. Sanches, I. A. Shelykh, and A. C. Seridonio,
Interaction induced hybridization of Majorana zero modes in a coupled quantum-dot–superconducting-nanowire hybrid system,
Phys. Rev. B \textbf{102}, 165104 (2020).

\bibitem{souto}  R. Seoane Souto, A. Tsintzis,  M. Leijnse, and J. Danon,
Probing Majorana localization in minimal Kitaev chains through a quantum dot,
Phys. Rev. Research \textbf{5}, 043182 (2023).

\bibitem{kenyi} R. Kenyi Takagui P\'{e}rez and A. A. Aligia,
Effect of interatomic repulsion on Majorana zero modes in a coupled quantum-dot–superconducting-nanowire hybrid system,
Phys. Rev. B \textbf{109}, 075416 (2024).


\bibitem{leij} M. Leijnse and K. Flensberg, Parity qubits and poor man’s
Majorana bound states in double quantum dots, Phys. Rev. B
\textbf{86}, 134528 (2012).

\bibitem{sau} J. D. Sau and S. D. Sarma, Realizing a robust practical Majorana chain in a quantum-dot-superconductor linear array, Nat.
Commun. \textbf{3}, 964 (2012).

\bibitem{liu} C.-X. Liu, G. Wang, T. Dvir, and M. Wimmer, Tunable super-
conducting coupling of quantum dots via Andreev bound states
in semiconductor-superconductor nanowires, Phys. Rev. Lett.
\textbf{129}, 267701 (2022).

\bibitem{tsin} A. Tsintzis, R. S. Souto, and M. Leijnse, Creating and detecting poor man’s Majorana bound states in interacting quantum dots,
Phys. Rev. B \textbf{106}, L201404 (2022).

\bibitem{bordin23} A. Bordin, G. Wang, C.-X. Liu, S. L. D. ten Haaf, N. van Loo, G. P. Mazur, D. Xu, D. van Driel, F. Zatelli, S. Gazibegovic, G.
Badawy, E. P. A. M. Bakkers, M. Wimmer, L. P. Kouwenhoven,
and T. Dvir,
Tunable crossed Andreev reflection and elastic
cotunneling in hybrid nanowires,
Phys. Rev. X \textbf{13}, 031031 (2023).

\bibitem{dvir} T. Dvir, G. Wang, N. van Loo, C.-X. Liu, G. P. Mazur,
A. Bordin, S. L. D. ten Haaf, S. L. D. ten Haaf, J.-
Y. Wang, D. van Driel, F. Zatelli, X. Li, F. K. Mali-
nowski, S. Gazibegovic, G. Badawy, E. P. A. M. Bakkers,
M. Wimmer, and L. P. Kouwenhoven,
Realization of a minimal Kitaev chain in coupled quantum dots,
Nature \textbf{614}, 445 (2023).

\bibitem{haaf} S. L. D. ten Haaf, Q. Wang, A. M. Bozkurt, C.-X. Liu,
I. Kulesh, P. Kim, D. Xiao, C. Thomas, M. J. Manfra,
T. Dvir, M. Wimmer, and S. Goswami,
A two-site Kitaev chain in a two-dimensional electron gas
Nature \textbf{630}, 329 (2024).

\bibitem{bordin2} Alberto Bordin, Xiang Li, David van Driel, Jan Cornelis Wolff, Qingzhen Wang, Sebastian L.D. ten Haaf, Guanzhong Wang, Nick van Loo, Leo P. Kouwenhoven, and Tom Dvir,
Crossed Andreev Reflection and Elastic Cotunneling in Three Quantum Dots Coupled by Superconductors,
Phys. Rev. Lett. \textbf{132}, 056602 (2024).

\bibitem{bordin3} Alberto Bordin, Chun-Xiao Liu, Tom Dvir, Francesco Zatelli, Sebastiaan L. D. ten Haaf, David van Driel, Guanzhong Wang, Nick van Loo, Thomas van Caekenberghe, Jan Cornelis Wolff, Yining Zhang, Ghada Badawy, Sasa Gazibegovic, Erik P.A.M. Bakkers, Michael Wimmer, Leo P. Kouwenhoven, and Grzegorz P. Mazur,
Signatures of Majorana protection in a three-site Kitaev chain,
arXiv:2402.19382


\bibitem{liu22} C.-X. Liu, G. Wang, T. Dvir, and M. Wimmer, Tun-
able Superconducting Coupling of Quantum Dots via
Andreev Bound States in Semiconductor-Superconductor
Nanowires, Phys. Rev. Lett. \textbf{129}, 267701 (2022).

\bibitem{tsin24} A. Tsintzis, R. S. Souto, K. Flensberg, J. Danon, and
M. Leijnse,
Majorana Qubits and Non-Abelian Physics
in Quantum Dot–Based Minimal Kitaev Chains,
PRX Quantum \textbf{5}, 010323 (2024).

\bibitem{pino} D. M. Pino, R. S. Souto, and R. Aguado,
Minimal Kitaev-transmon qubit based on double quantum dots,
Phys. Rev. B \textbf{109}, 075101 (2024).

\bibitem{liu24} Zhi-Hai Liu, Chuanchang Zeng, H. Q. Xu,
Coupling of quantum-dot states via elastic-cotunneling and crossed Andreev reflection in a minimal Kitaev chain,
Phys. Rev. B \textbf{110}, 115302 (2024).

\bibitem{luet} Melina Luethi, Henry F. Legg, Daniel Loss, and Jelena Klinovaja,
From perfect to imperfect poor man's Majoranas in minimal Kitaev chains,
Phys. Rev. B \textbf{110}, 245412 (2024).

\bibitem{sanches} J.E. Sanches, L.T. Lustosa, L.S. Ricco, H. Sigurosson,
M. de Souza, M.S. Figueira, E. Marinho Jr., and A.C. Seridonio,
Spin-exchange induced spillover on poor man's Majoranas in minimal Kitaev chains,
J. Phys.: Condens. Matter \textbf{37}, 205601 (2025).

\bibitem{pers} F. Persson, C.M. Wilson, M. Sandberg, P. Delsing,
Fast readout of a single Cooper-pair box using its quantum capacitance,
Phys. Rev. B \textbf{82}, 134533 (2010).

\bibitem{gino} Eran Ginossar and Eytan Grosfeld,
Microwave transitions as a signature of coherent parity mixing effects in the
Majorana-transmon qubit,
Nat. Commun. \textbf{5}, 4772 (2014).

\bibitem{vaey} Jukka I. V\"{a}yrynen, Gianluca Rastelli, Wolfgang Belzig, and Leonid I. Glazman,
Microwave signatures of Majorana states in a topological
Josephson junction,
Phys. Rev. B \textbf{92}, 134508 (2015).

\bibitem{avila} J. \'Avila, Prada, P. San-Jose, and R. Aguado,
Majorana oscillations and parity crossings in semiconductor nanowire-based transmon qubits,
Phys. Rev. Research \textbf{2}, 033493 (2020).

\bibitem{larsen} T. W. Larsen, K. D. Petersson, F. Kuemmeth, T. S. Jespersen,
P. Krogstrup, J. Nygard and C. M. Marcus,
Semiconductor-Nanowire-Based Superconducting Qubit,
Phys. Rev. Lett. \textbf{115}, 127001 (2015).

\bibitem{caspa} Lucas Casparis, Malcolm R. Connolly, Morten Kjaergaard, Natalie J. Pearson, Anders  Kringh\o j, Thorvald W. Larsen, Ferdinand Kuemmeth, Tiantian Wang, Candice Thomas, Sergei Gronin, Geoffrey C. Gardner, Michael J. Manfra, Charles M. Marcus, Karl D. Petersson,
Superconducting Gatemon Qubit based on a Proximitized Two-Dimensional Electron Gas,
Nature Nanotechnology \textbf{13}, 915 (2018).

\bibitem{barge20} A. Bargerbos, W. Uilhoorn, C.-K. Yang, P. Krogstrup,
L. P. Kouwenhoven, G. de Lange, B. van Heck, and A. Kou,
Observation of vanishing charge dispersion of a nearly open
superconducting island,
Phys. Rev. Lett. \textbf{124}, 246802 (2020).

\bibitem{krin} A. Kringh\o j, B. van Heck, T. W. Larsen, O. Erlandsson,
D. Sabonis, P. Krogstrup, L. Casparis, K. D. Petersson, and C. M.
Marcus,
Suppressed charge dispersion via resonant tunneling
in a single-channel transmon,
Phys. Rev. Lett. \textbf{124}, 246803 (2020).

\bibitem{barge22} A. Bargerbos, M. Pita-Vidal, R. \v{Z}itko, J.\'{A}vila,
L. J. Splitthoff,
L. Gr\"{u}nhaupt, J. J. Wesdorp, C. K. Andersen, Y. Liu,
L. P. Kouwenhoven, R. Aguado, A. Kou, and B. van Heck,
Singlet-doublet transitions of a quantum dot Josephson junction
detected in a transmon circuit,
PRX Quantum \textbf{3}, 030311 (2022).

\bibitem{pita} Marta Pita-Vidal, Arno Bargerbos, Rok \v{Z}itko, Lukas J. Splitthoff, Lukas Gr\"{u}nhaupt, Jaap J. Wesdorp, Yu Liu, Leo P. Kouwenhoven, Ram\'on Aguado, Bernard van Heck, Angela Kou, and Christian Kraglund Andersen,
Direct manipulation of a superconducting spin qubit strongly coupled to a transmon qubit,
Nature Physics \textbf{19}, 1110 (2023).

\bibitem{gyenis} Andr\'as Gyenis, Agustin Di Paolo, Jens Koch, Alexandre Blais,
Andrew A. Houck, and David I. Schuster,
Moving beyond the Transmon: Noise-Protected Superconducting Quantum Circuits,
PRX Quantum \textbf{2}, 030101 (2021).

\bibitem{matu} F.J. Matute-Ca\~{n}adas, L. Tosi, and A. Levy Yeyati,
Quantum Circuits with Multiterminal Josephson-Andreev Junctions,
PRX Quantum \textbf{5}, 020340 (2024).

\bibitem{torres} Juan Daniel Torres Luna, A. Mert Bozkurt, Michael Wimmer,
and Chun-Xiao Liu,
Flux-tunable Kitaev chain in a quantum dot array,
SciPost Phys. Core \textbf{7}, 065 (2024).

\bibitem{smith} Thomas B. Smith, Maja C. Cassidy , David J. Reilly,
Stephen D. Bartlett, and Arne L. Grimsmo,
Dispersive Readout of Majorana Qubits,
PRX Quantum \textbf{1}, 020313 (2020).

\bibitem{Ganga} Suhas Gangadharaiah, Bernd Braunecker, Pascal Simon,
and Daniel Loss,
Majorana Edge States in Interacting One-Dimensional Systems,
Phys. Rev. Lett. \textbf{107}, 036801 (2011).

\bibitem{tho13} R. Thomale, S. Rachel, and P. Schmitteckert,
Tunneling spectra simulation of interacting Majorana wires,
Phys. Rev. B \textbf{88}, 161103(R) (2013).

\bibitem{Pandey2} Bradraj Pandey , Narayan Mohanta, and Elbio Dagotto,
Out-of-equilibrium Majorana zero modes in interacting Kitaev chains,
Phys. Rev. B \textbf{107},  L060304 (2023).

\bibitem{miao17} Jian-Jian Miao, Hui-Ke Jin, Fu-Chun Zhang, and Yi Zhou,
Majorana zero modes and long range edge correlation in
interacting Kitaev chains: analytic solutions and density-matrix-
renormalization-group study,
Scientific Reports \textbf{8}, 488 (2018).

\bibitem{ger16} N. M. Gergs, L. Fritz, and D. Schuricht,
Topological order in the Kitaev/Majorana chain in the presence of disorder and interactions,
Phys. Rev. B \textbf{93}, 075129 (2016).

\bibitem{camja} A. Camjayi,  L. Arrachea, A. Aligia and F. von Oppen,
Fractional spin and Josephson effect in time-reversal-invariant topological superconductors,
Phys. Rev. Lett. \textbf{119}, 046801 (2017).

\bibitem{wiec} A. Wieckowski and A. Ptok, 
Influence of long-range interaction on Majorana zero modes,
Phys. Rev. B \textbf{100}, 144510 (2019).

\bibitem{pandey} B. Pandey, N. Kaushal, G. Alvarez, and  E, Dagotto,
Majorana zero modes in Y-shape interacting Kitaev wires,
npj Quantum Materials \textbf{8}, 51 (2023).

\bibitem{son} J. H. Son, J. Alicea, and O. I. Motrunich, Edge states
of two-dimensional time-reversal invariant topological
superconductors with strong interactions and disorder: A view
from the lattice, Phys. Rev. B \textbf{109}, 035138 (2024).

\bibitem{samue} William Samuelson, Viktor Svensson, and Martin Leijnse,
Minimal quantum dot based Kitaev chain with only local superconducting proximity effect,
Phys. Rev. B \textbf{109}, 035415 (2024).

\bibitem{chine} L. M. Chinellato, C. J. Gazza, A. M. Lobos, and A. A. Aligia,
Topological phases of strongly interacting time-reversal invariant topological superconducting chains under a magnetic field,
Phys. Rev. B \textbf{109}, 064503 (2024).

\bibitem{souto25} Rub\'en Seoane Souto, Virgil V. Baran, Maximilian Nitsch, Lorenzo Maffi,
Jens Paaske, Martin Leijnse, and Michele Burrello,
Majorana modes in quantum dots coupled via a floating superconducting island,
Phys. Rev. B \textbf{111}, 174501 (2025).

\bibitem{rafa1} R. S\'anchez and M. B\"uttiker,
Optimal energy quanta to current conversion,
Phys. Rev. B \textbf{83}, 085428 (2011).

\bibitem{rafa2} R. S\'anchez and M. B\"uttiker,
Detection of single-electron heat transfer statistics,
EPL \textbf{100}, 47008 (2012).

\bibitem{ruoko} T. Ruokola and T. Ojanen,
Single-electron heat diode: Asymmetric heat transport between electronic
reservoirs through Coulomb islands,
Phys. Rev. B \textbf{83}, 241404(R) (2011).

\bibitem{yada} H. K. Yadalam and U. Harbola,
Statistics of heat transport across a capacitively coupled double quantum dot circuit,
Phys. Rev. B \textbf{99}, 195449 (2019).

\bibitem{heat} A. A. Aligia, D. P\'erez Daroca, L. Arrachea,
and P. Roura-Bas,
Heat current across a capacitively coupled double quantum dot
Phys. Rev. B \textbf{101}, 075417 (2020).

\bibitem{dar1} D. P\'erez Daroca, P. Roura-Bas and A. A. Aligia,
Thermoelectric properties of a double quantum dot out of equilibrium in Kondo and intermediate valence regimes,
Phys. Rev. B \textbf{108}, 155117 (2023).

\bibitem{dar2} D. P\'erez Daroca, P. Roura-Bas and A. A. Aligia,
Role of asymmetry in thermoelectric properties of a double quantum dot out of equilibrium,
Phys. Rev. B \textbf{111}, 045134 (2025).

\bibitem{ferrel}  R. A. Ferrell and R. E. Prange,
Self-Field Limiting of Josephson Tunneling of Superconducting Electron Pairs,
Phys. Rev. Lett. \textbf{10}, 479 (1963).

\bibitem{newr} R. S. Newrock, C. J. Lobb, U. Geigenm\"uller, and M. Octavio,
The two-dimensional physics of Josephson junction arrays,
Solid State Physics \textbf{54}, 263 (2000).

\bibitem{note}  D. M. Pino, private communication.

\bibitem{Huang} Tianyu Huang, Rui Zhang, Xiaopeng Li, Xiong-Jun Liu, X. C. Xie, Yijia Wu,
Unified model for non-Abelian braiding of Majorana and Dirac fermion zero modes,
arXiv:2410.05957


\end{thebibliography}
\end{document}